\documentclass[10pt,conference]{IEEEtran}

\usepackage{lgrind}
\usepackage{epsfig}
\usepackage{amsmath}    
\usepackage{graphicx}
\usepackage{amssymb}
\pdfoutput=1

\begin{document}

\title{PPM demodulation: On approaching fundamental limits of optical communications}

\author{
\authorblockN{Saikat Guha, {\em{Member IEEE}}}
\authorblockA{Raytheon BBN Technologies \\
Cambridge, MA 02138, USA \\
sguha@bbn.com}
\and
\authorblockN{Jonathan L. Habif}
\authorblockA{Raytheon BBN Technologies \\
Cambridge, MA 02138, USA \\
jhabif@bbn.com}
\and
\authorblockN{Masahiro Takeoka}
\authorblockA{
National Inst. of Inf. and Comm. Technology \\
Tokyo 184-8795, Japan \\
takeoka@nict.go.jp}
}

\maketitle

\begin{abstract}
We consider the problem of demodulating $M$-ary optical PPM (pulse-position modulation) waveforms, and propose a structured receiver whose mean probability of symbol error is smaller than all known receivers, and approaches the quantum limit. The receiver uses photodetection coupled with optimized phase-coherent optical feedback control and a phase-sensitive parametric amplifier. We present a general framework of optical receivers known as the {\em conditional pulse nulling receiver}, and present new results on ultimate limits and achievable regions of spectral versus photon efficiency tradeoffs for the single-spatial-mode pure-loss optical communication channel. 
\end{abstract}
\vspace{-5pt}
\section{Communications under the Poisson regime}

As per the semi-classical theory of photodetection, the output of a photodetector (e.g., a PIN diode) is the superposition of two statistically independent Poisson point processes: one with a constant rate $\lambda_d$ ({\em dark-current noise}) and another whose time-varying rate $\lambda(t)$ is equal to the instantaneous squared-magnitude intensity of the $\sqrt{{\text {photons/m}}^2{\rm sec}}$ unit optical signal field integrated over the receiver aperture. A single-photon detector (SPD) is a low-noise photodetection device that only has single-photon resolving capability, i.e., it registers photon clicks with a Poisson rate $\lambda(t)$, but after each click, the detector is inactive for a minimum {\em dead time} before it can click again, whether or not any light impinges on its active surface during that time duration (see Fig.~\ref{fig:PPMfigure}). Two seminal papers in the late sixties laid the foundations of optical communication theory as an elegant counterpart of the popular model of the additive white Gaussian noise channel. The first was a treatise by Bar-David~\cite{Bar1969}, who worked out explicitly the {\em Poisson matched filter} for the optimal detection of a set of independent arbitrarily time-varying Poisson point processes of rates $\lambda_k(t)$, $1 \le k \le M$, which established a conceptual framework for the optimal estimation theory when laser-light modulated waveforms are detected by a photodetector. The other was a paper by Helstrom~\cite{Hel1967}, which established a concrete groundwork of finding the ultimate limits on optical demodulation as required by the laws of quantum mechanics. 

In recent years, it has been shown that the ultimate classical-information capacity of the optical channel with linear loss and background thermal noise can be attained by single-use coherent-state encoding, i.e., non-classical states of light are not needed to achieve the ultimate channel capacity~\cite{Gio2004, Sha2005}. Nevertheless, conventional receivers such as direct detection and coherent detection fall short of achieving the ultimate capacity limits, and optimal {\em joint detection receivers}, which make joint quantum-optical measurements over long codeword blocks (which may include receivers that use adaptive measurements and classical feedforward), may be needed. Consider a collection of $M$ coherent-state temporal waveforms $\psi_k(t), 1 \le k \le M$ forming a set of modulation symbols, each of duration $T$ seconds. Let $\psi_k(t) = {\rm Re}\left[\Psi_k(t)e^{-j(\omega_0t+\phi_k)}\right]$ be the $\sqrt{\text{photons/sec}}$ quasi-monochromatic passband optical field at the receiver when the $k^{\rm th}$ modulation symbol is transmitted, where $\Psi_k(t)$ is the baseband received temporal pulse shape, $\omega_0$ the center frequency of transmission, and $\phi_k$ is a modulation phase (non-zero for quadrature-phase modulation sets). For any set of $M$ coherent-state signals, the minimum mean probability of error in distinguishing the $M$ signals---as permitted by quantum mechanics---can be calculated using the Helstrom bound~\cite{Hel1976}. Nevertheless, the structured optical receiver that achieves this minimum probability of error is not known in general except for the special case of binary modulation ($M=2$). A structured realization of the quantum-optimal receiver for binary modulation (such as BPSK or OOK) was derived in theory by Dolinar in 1973~\cite{Dol1973} and the first preliminary implementation of the concept was demonstrated recently, more than four decades after its invention~\cite{Coo2007}. The receiver uses photodetection coupled with optimal time-varying phase-coherent feedback in the $[0,T]$ symbol interval. Near-quantum-optimal structured receivers have been proposed for binary modulation~\cite{Ken1973}, pulse-position modulation (PPM)~\cite{Dol1983} and phase-shift keying (PSK)~\cite{ Bon1993}. In this paper, we combine all the optimal and near-optimal optical demodulation receivers that have been found to date under one general class of structured receivers---the {\em{conditional pulse nulling}} (CPN) receiver---which we then use to obtain the best structured PPM receiver known to date. Moreover, it is highly likely that the CPN strategy, along with optimal feedback control, may have the capability to attain the quantum-limited minimum error rate for any coherent-state modulation format. Here we will concentrate on PPM demodulation to illustrate the ideas. High-sensitivity PPM receivers could be key to developing high-speed deep-space communication systems~\cite{Bor2004}. We will purposely refrain from using explicit quantum notation and use a semiclassical analysis wherever possible. When a full quantum analysis is a must, we will state the physical results without explicit derivations.

\section{Pulse position modulation: Direct detection receiver and the Helstrom bound}

\begin{figure}[h]
\begin{center}
\includegraphics[width=0.7\columnwidth,angle=0]{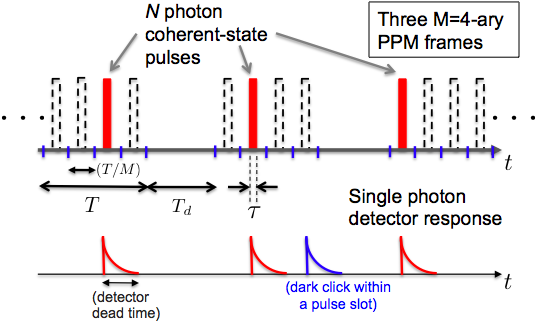}
\end{center}
\vspace{-10pt}
\caption{Pulse position modulation (PPM).}
\label{fig:PPMfigure}
\end{figure}

$M$-ary PPM encodes digital data by positioning a $\tau$-second-long flat-topped laser pulse in one of $M$ equal-duration $T/M$-second intervals of a $T$-second long PPM modulation symbol (see Fig.~\ref{fig:PPMfigure}). Each symbol may be separated from the next by $T_d \ge 0$ seconds. The task of the receiver is to discriminate between the $M$ possible received PPM symbols with the minimum probability of error. Let the mean photon number in a received PPM pulse be $N$. Hence, if a pulse bearing slot is detected using a photon-counting (direct-detection) receiver of quantum efficiency $\eta < 1$, the probability that no clicks are registered is given by $e^{-(\eta{N}+\lambda_d\tau)}$, where $\lambda_d$ is the dark photon arrival rate and $P_d \equiv 1-e^{-\lambda_d\tau}$ is the dark-click probability of the detector gated to a pulse slot.

The conventional PPM receiver photodetects each of the $M$ pulse slots sequentially and declares the slot that produces the maximum photon counts as the pulse-bearing slot. If two or more slots produce the same number of photon counts, the receiver declares one of those slots---chosen uniformly randomly---as the pulse-bearing slot. For ideal signal-shot-noise limited detector operation (i.e., in the absence of background, dark noise, and thermal (Johnson) noise), a single-photon detector (SPD) has the same performance as what is obtained by a full photon-number-resolving detector. Assuming the PPM symbols to be equally likely, the shot-noise-limited mean symbol error probability achieved by the unity quantum-efficiency direct-detection (DD) receiver is given by:
\begin{equation}
P_{e,{\rm DD}} = \frac{M-1}{M}e^{-N}.
\label{eq:DDerror}
\end{equation}
Helstrom showed that the ultimate limit to the minimum probability of error (MPE) of discriminating the PPM symbols, as required by quantum mechanics, is given by~\cite{Hel1976}:
\begin{equation}
P_{e,\min} = \frac{M-1}{M^2}\left[\sqrt{1+(M-1)e^{-N}}-\sqrt{1-e^{-N}}\right]^2,
\end{equation}
which has a $3$-dB higher error exponent than the error rate of the DD receiver~\eqref{eq:DDerror} in the high photon number regime, i.e., $P_{e,\min} \sim e^{-2N}$, when $Me^{-N} \ll 1$. Fig.~\ref{fig:enhancedreceivererrorrates} plots the error rates as a function of $N$ for the DD and the optimal receivers. The intuitive reason why DD falls short of achieving the ultimate performance limit is, it assembles information from each pulse slot individually, without an effective overall strategy.

\section{The conditional pulse nulling (CPN) receiver}

\begin{figure}
\begin{center}
\includegraphics[width=8cm,angle=0]{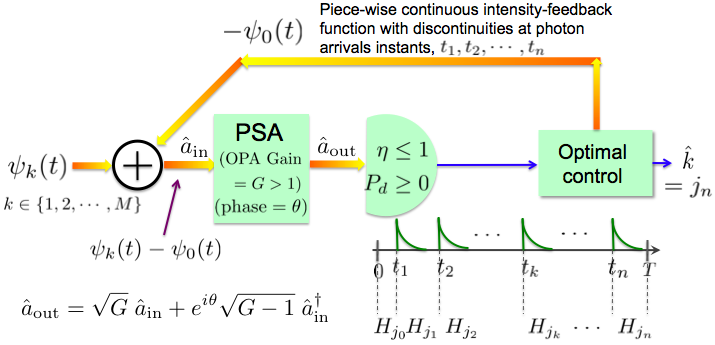}
\end{center}
\vspace{-10pt}
\caption{A schematic of the general conditional pulse nulling (CPN) receiver.}
\label{fig:generalCPNschematic}
\end{figure}

\subsection{The general CPN receiver}
The CPN receiver is shown schematically in Fig.~\ref{fig:generalCPNschematic}. $\psi_k(t)$ is the $T$-sec-long received pulse, and the receiver's task is to generate the MPE estimate $\hat k$ of the transmitted modulation symbol $k$, $1 \le k \le M$. The receiver begins at $t=0$ with a symbol $j_0$ as its initial hypothesis $H_{j_0}$ (i.e., ``$\psi_{j_0}(t)$ is the received symbol waveform"). An optimal control system generates an optical feedback field $-\psi_0(t)$, $t \in 0^+$, which is added coherently to the incoming pulse on an asymmetric beamsplitter. The sum field $\psi_k(t) - \psi_0(t)$ is amplified by a phase-sensitive amplifier (PSA) of gain $G$. The PSA is a quantum-optical device also known as a {\em squeezer} or a {\em noiseless amplifier}, which produces a squeezed-coherent state for a coherent-state input\footnote{A squeezed-coherent state is a quantum state of light, whose photodetection statistics cannot be described quantitatively correctly using a semiclassical analysis alone. It is a minimum-uncertainty product state; i.e., the quantum noise variance of one of the field quadratures is {\em{squeezed}} whereas the noise-variance of the other orthogonal field quadrature is {\em{amplified}} proportionately, from the coherent-state quadrature-noise variance levels~\cite{Sha1985}.}. The displacement and the squeezing operations together exhaust all so-called {\em{Gaussian unitary operations}}. The output of the PSA is detected by a SPD. In the absence of the PSA (i.e., $G=1$) and assuming zero dark current ($\lambda_d=0$), the SPD registers clicks at a Poisson arrival rate $\lambda(t) = \eta|\psi_k(t) - \psi_0(t)|^2$, where $\eta$ is the quantum efficiency of the detector. With PSA gain $G>1$ and phase $\theta = 0$, the full photon-statistics of the phase-sensitively amplified field is a far more complicated non-Poissonian distribution. But the only physically relevant quantity for the SPD operation is the probability that a click is registered or not. With sub-unity SPD quantum efficiency $\eta$ and non-zero dark photon arrival rate $\lambda_d$, the probability of registering a click over a time interval $[t_1, t_2] \in [0,T]$ is given by ${\rm{Pr_{\rm click}(t_1,t_2)}} = 1 - p_0(t_1,t_2)$, where [derivation omitted],
\begin{eqnarray}
p_0(t_1,t_2) &=& \frac{1-\lambda_d|t_2-t_1|}{\sqrt{G-(1-\eta)^2(G-1)}} \times \nonumber\\
&&\exp\left[-\frac{\eta\left(\sqrt{G}+\sqrt{G-1}\right)^2N(t_1,t_2)}{1+\eta\sqrt{G-1}\left(\sqrt{G}+\sqrt{G-1}\right)}\right],
\label{eq:probzeroclick}
\end{eqnarray}
where $N(t_1,t_2) = \int_{t_1}^{t_2}|\psi_k(\tau) - \psi_0(\tau)|^2{{\rm d}\tau}$ is the integrated mean photon number in the nulled waveform in $[t_1,t_2]$. At the $k^{\rm th}$ SPD click ($1 \le k \le n$), the receiver switches it's current guess from symbol-$j_{k-1}$ to symbol-$j_k$ as per an optimal control algorithm that attempts to minimize the evolving Bayesian estimate of the mean decoding error probability, based on its knowledge of all the past arrival times $\left\{t_1, \cdots, t_{k}\right\}$. The receiver's hypothesis at $t=T$ seconds is declared as the output, i.e., ${\hat k} = j_n$. 

\subsection{Sequential exact-nulling CPN receiver}
\begin{figure}
\begin{center}
\includegraphics[width=0.7\columnwidth,angle=0]{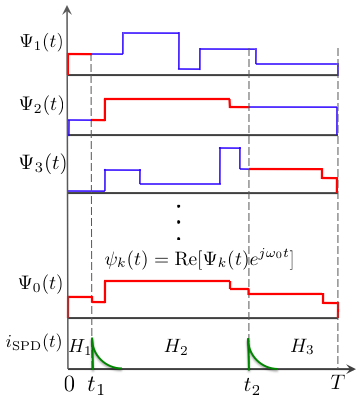}
\end{center}
\vspace{-10pt}
\caption{Sequential exact-nulling CPN receiver --- shown for an amplitude-modulation symbol set. In this example, all the received modulation symbols $\psi_k(t)$, $1 \le k \le M$ have the same carrier frequency and phase, but have different real baseband temporal shapes $\Psi_k(t)$. The receiver starts with hypothesis $H_1$ (i.e., ``$\psi_1(t)$ is received") at $t=0$ and on each click registered by the SPD, it advances its hypothesis by one. In this depiction, the current hypothesis at $t=T$ is $H_3$, hence $\psi_3(t)$ is declared as the final estimate for the received symbol (${\hat k} = 3$). $i_{\rm SPD}(t)$ is the photocurrent output of the single photon detector.}
\label{fig:generalnulling}
\end{figure}

We consider now a special case of the general CPN receiver. Let the order of the hypotheses be chosen as $H_1, \cdots, H_M$. When the current hypothesis is $H_k$, the feedback function is chosen as $\phi_0(t) = \phi_k(t)$, i.e., a phase-inverted amplitude-matched {\em{nulling}} pulse corresponding to the current-guess modulation waveform is applied to the incoming received symbol waveform, such that if the current guess is indeed the true transmitted symbol, in the limit of $\lambda_d \to 0$, no clicks will be registered with probability $1$. Thus in the limit of zero dark current and perfect nulling, just a single detector click confirms against the current hypothesis with certainty. An illustration of this special case, drawn only for an amplitude-modulated symbol set for clarity of illustration, is shown in Fig.~\ref{fig:generalnulling}. 

The first designs of structured near-quantum optimal receivers were proposed by Robert Kennedy~\cite{Ken1973} for binary modulation, Sam Dolinar~\cite{Dol1983} for $M$-ary PPM modulation, and by Roy Bondurant~\cite{Bon1993} for $M$-ary PSK modulation --- {\em{all}} of which are in fact the sequential exact-nulling CPN receiver specialized for the respective modulation sets. Takeoka et. al.'s generalized Kennedy receiver for near-optimal binary demodulation~\cite{Tak2008} is a simple extension of the sequential exact-nulling CPN receiver, where the feedback function $\phi_0(t) = \phi_k(t) + \beta$ has an optimized constant amplitude-offset $\beta > 0$ with the actual symbol hypothesis. The conventional receiver to demodulate $M$-ary PSK signals is heterodyne detection, whereas the conventional receiver to demodulate PPM signals is direct detection. In both of these cases, the respective CPN receiver's performance surpasses that of the conventional receiver and have the same error-exponent as the quantum-optimal receiver in the limit of high photon numbers, although the CPN receiver's performance approaches that of the conventional receiver in the low photon number regime (see Fig.~\ref{fig:enhancedreceivererrorrates}). Furthermore, in each case the quantum-limited minimum symbol error rate has a $3$-dB higher error exponent than the respective conventional receiver in the high photon number regime. Section~\ref{sec:CPNadvanced} will show two strategies to bridge the gap between the PPM symbol error rates of the CPN and the quantum-optimal receivers (the green and the blue curves in Fig.~\ref{fig:enhancedreceivererrorrates}) in the low-photon number regime. 

\subsection{The baseline (sequential exact-nulling) CPN receiver for PPM demodulation}
\begin{figure}
\begin{center}
\includegraphics[width=0.8\columnwidth,angle=0]{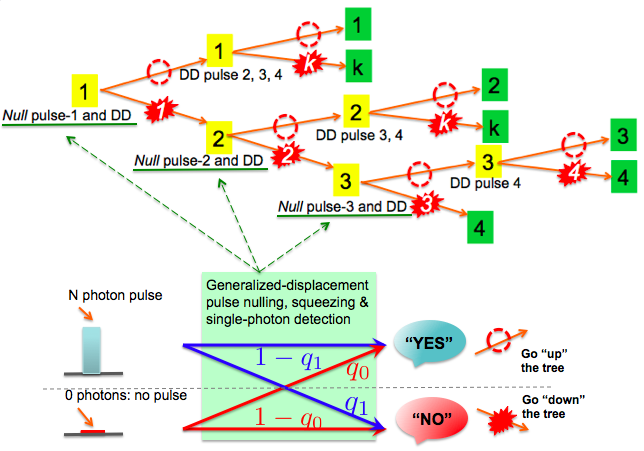}
\end{center}
\caption{Decision tree of the conditional pulse nulling (CPN) receiver for PPM demodulation. Also are shown the transition probabilities at the root nodes of each sub tree when a generalized CPN receiver (Type 1 or Type 2) is used. The values of the transition probabilities are given in Section~\ref{sec:CPNadvanced}.}
\label{fig:cpntree}
\end{figure}

The baseline CPN receiver for PPM demodulation~\cite{Dol1983} is a specialization of the sequential exact-nulling CPN receiver to the PPM alphabet, which reduces to a sequential pulse-nulling algorithm depicted schematically by a decision tree in Fig.~\ref{fig:cpntree}. 
The receiver starts with hypothesis $H_1$ (i.e., the pulse is in slot $1$). It applies a $N$-mean-photon phase-inverted coherent-state feedback pulse to the first slot ($[0,T/M]$) that would null the pulse completely if it were indeed present in slot $1$. The receiver then direct-detects the interval $[0,T/M]$ using a SPD. If no click is registered, the receiver simply detects each of the remaining $M-1$ slots using a SPD (with no feedback applied) while it continues to believe in hypothesis $H_1$, unless any one of those $M-1$ slots $j$ produces a click that would confirm hypothesis $H_j$. If a click is registered in the first slot, then hypothesis $H_1$ is ruled out. Thus the problem reduces to an $(M-1)$-ary version of the $M$-ary detection problem with which we began (see Fig.~\ref{fig:cpntree}). Note that if any one of the hypotheses $H_j$, $2 \le j \le M$ were true, for a symbol error to occur, it must escape detection in both the first interval $[0,T/M]$, and its {\em own} interval $[(j-1)T/M, jT/M]$, accruing an error-exponent of $2N$ along this route---roughly speaking. An error analysis of this receiver (see following sub-section) shows that the mean symbol error rate achievable is given by,
\begin{equation}
P_{e,{\rm CPN}} = \frac{1}{M}\left[(1-e^{-N})^M + Me^{-N}-1\right].
\label{eq:cpnerror}
\end{equation}
For high information rates and low signal energies (i.e., $Me^{-N} \gg 1$), the above expression yields the same error-exponent as the standard DD receiver (i.e. $P_e \sim e^{-N}$). However, for $Me^{-N} \ll 1$, the baseline CPN receiver attains the error exponent of the optimal MPE receiver (i.e., $P_e \sim e^{-2N}$). Fig.~\ref{fig:enhancedreceivererrorrates} shows the comparative performance of all three receivers.

\subsection{Enhanced CPN receiver architectures}
\label{sec:CPNadvanced}

\noindent {\bf Type 1: Optimal-displacement pulse nulling} --- The baseline CPN receiver uses a feedback function $\phi_0(t) = \phi_k(t)$ when the current guess is symbol-$k$, such that the feedback, ideally, exactly nulls the modulation symbol. We will now allow for (and optimize over) a constant offset $\beta$ in the feedback pulse, i.e., $\phi_0(t) = \phi_k(t) + \beta$ with $\beta > 0$, such that the feedback pulse has slightly more photons than the actual pulse. Let us say that the received PPM pulses have $N$ mean photons, and the phase-inverted nulling pulses have $n_1 > N$ mean photons. Then, if the pulse is present, the nulled slot will be phase-inverted with respect to the original pulse and will have $n_0$ mean photons, where $n_1 = \left(\sqrt{n_0}+\sqrt{N}\right)^2$. If a slot is actually empty then a nulled slot will contain a phase-inverted $n_1$-mean-photon pulse. The {\em Type 1} CPN receiver uses optimal-displacement pulse nulling (i.e., an optimal value of $n_0$) whenever it needs to null a slot based on the decision tree as shown in Fig.~\ref{fig:cpntree}.

\noindent {\bf Type 2: Phase-sensitive amplification (squeezing)} --- The {\em Type 2} receiver is identical to the {\em Type 1} receiver except that a phase-sensitive amplifier (PSA) is used to squeeze the nulled pulse by an optimal amount, as shown in Fig.~\ref{fig:generalCPNschematic}. We have found analytically that the optimum PSA phase is $\theta = 0$, but we numerically optimize over the PSA gain $G \ge 1$, as well as the nulling photon-number residue $n_0 \ge 0$.

\subsubsection*{Error analysis of the CPN receiver}
In terms of the decision tree as shown in Fig.~\ref{fig:cpntree}, the only difference between the enhanced versions of the CPN receiver and the baseline CPN receiver are the transition probabilities at the root node of each subtree on the lower branches of the main tree (shown using green underscore in Fig.~\ref{fig:cpntree}). The transition probabilities are shown in the bottom part of Fig.~\ref{fig:cpntree}. At the root node, the probability of going {\em up} the tree is the probability that no click was registered, and the probability of going {\em down} the tree is the probability of a click. The no-click probability given there was no pulse in the slot is denoted as $q_0$ and the probability of a click given there was a pulse in the slot, is denoted $q_1$. Thus, $q_0$ and $1-q_1$, for the Type 2 receiver are given by Eq.~\eqref{eq:probzeroclick} with $\lambda_d|t_2-t_1| \equiv P_d$ being the gated single slot dark-click probability and $N(t_1,t_2)$ been substituted by $n_1 \equiv (\sqrt{N}+\sqrt{n_0})^2$ and $n_0$, respectively. Substituting $G=1$ reduces the expressions to those for the Type 1 CPN receiver, and further substituting $n_0=0$ reduces the expressions to those for the baseline CPN receiver, for which $q_0 = (1-P_d)e^{-\eta{N}}$ and $q_1=P_d$ hold. 

\begin{figure}
\begin{center}
\includegraphics[width=0.8\columnwidth,angle=0]{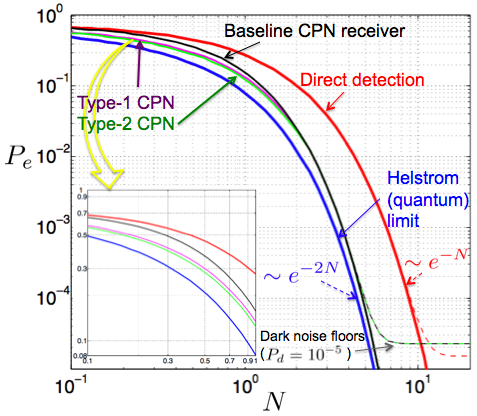}
\end{center}
\caption{Quaternary PPM symbol error rate (assuming $\eta=1$ and $\lambda_d = 0$), for the direct-detection receiver (red), the baseline CPN receiver (green), the quantum MPE limit (blue), the Type 1 (magenta) and the Type 2 (green) CPN receivers. The respective dark-noise {\em{floors}} for the direct-detection and the baseline CPN receivers are also shown for per-slot dark-click probability $P_d \equiv \lambda_d\tau = 10^{-5}$. The inset shows a zoomed-in plot of the same mean symbol error rates, showing the distinct advantage of the Type 2 CPN receiver over the baseline CPN receiver and direct-detection.}
\label{fig:enhancedreceivererrorrates}
\end{figure}

Let $P_M$ be the mean symbol error probability and assume equally likely PPM symbols. Then, we have the recursion:
\begin{eqnarray}
P_M &=& \frac{1}{M}\left[(1-q_1)\left(1-(1-P_d)^{M-1}\right)+q_1\right] + \frac{M-1}{M} \times \nonumber \\
&&\left[q_0\left(1-\frac{\left(1-(1-P_d)e^{-\eta{N}}\right)(1-(1-P_d)^{M-1})}{(M-1)P_d}\right)\right. \nonumber \\
&&\left.+(1-q_0)P_{M-1}\right],
\label{eq:recursionPdnonzero}
\end{eqnarray}
whose solution is given by:
\begin{eqnarray}
P_M &=& \frac1M\left[\frac{A(1-B^{M-1})}{(1-B)}+\frac{\mu{D}(\mu^{M-1}-B^{M-1})}{(\mu-B)}\right. \nonumber \\
&&\left.+\frac{(M-1)C-MCB+CB^M}{(1-B)^2}\right],
\end{eqnarray}
where $A = 1 - (q_0/P_d)(1-(1-P_d)e^{-{\eta}N})$, $D=q_1-A$, $B=1-q_0$, $C=q_0$ and $\mu = 1-P_d$. For $P_d = 0$, taking the limit of Eq.~\eqref{eq:recursionPdnonzero} as $P_d \to 0$ and solving the recursion yields:
\begin{eqnarray}
P_M &=& \frac1M\left[\frac{D^\prime(1-B^{{\prime}{M-1}})}{(1-B^\prime)}\right.\nonumber \\
&&\left.+\frac{(M-1)C^\prime-MC^{\prime}B^{\prime}+C^{\prime}B^{{\prime}M}}{(1-B^{\prime})^2}\right],
\label{eq:PM_Pdzero}
\end{eqnarray}
where  $B^\prime=1-q_0$, $C^\prime=q_0e^{-{\eta}N}$ and $D^\prime=q_1$. It is easy to verify that for the baseline CPN receiver with $\eta=1$ and $P_d=0$, (i.e., $q_0 = e^{-{N}}$, $q_1=0$), Eq.~\eqref{eq:PM_Pdzero} reduces to Eq.~\eqref{eq:cpnerror}. Fig.~\ref{fig:enhancedreceivererrorrates} shows the mean symbol error rates for demodulating $4$-ary PPM symbols. It shows that there is a distinct advantage of the Type 1 and the Type 2 CPN receiver architectures over the direct-detection limit at low photon numbers. The optimal nulling amplitudes in both the Type 1 and 2 CPN receivers are likely to be higher in the initial slots and decrease as we go {\em down} the decision tree, contrary to our assumption of a constant nulling amplitude. Furthermore, there is likely to be additional performance improvement when the PSA gain and phase is optimally updated sequentially along the pulse slots. 


\subsection{Channel capacity of PPM and the ultimate Holevo limit}
\begin{figure}
\begin{center}
\includegraphics[width=0.8\columnwidth,angle=0]{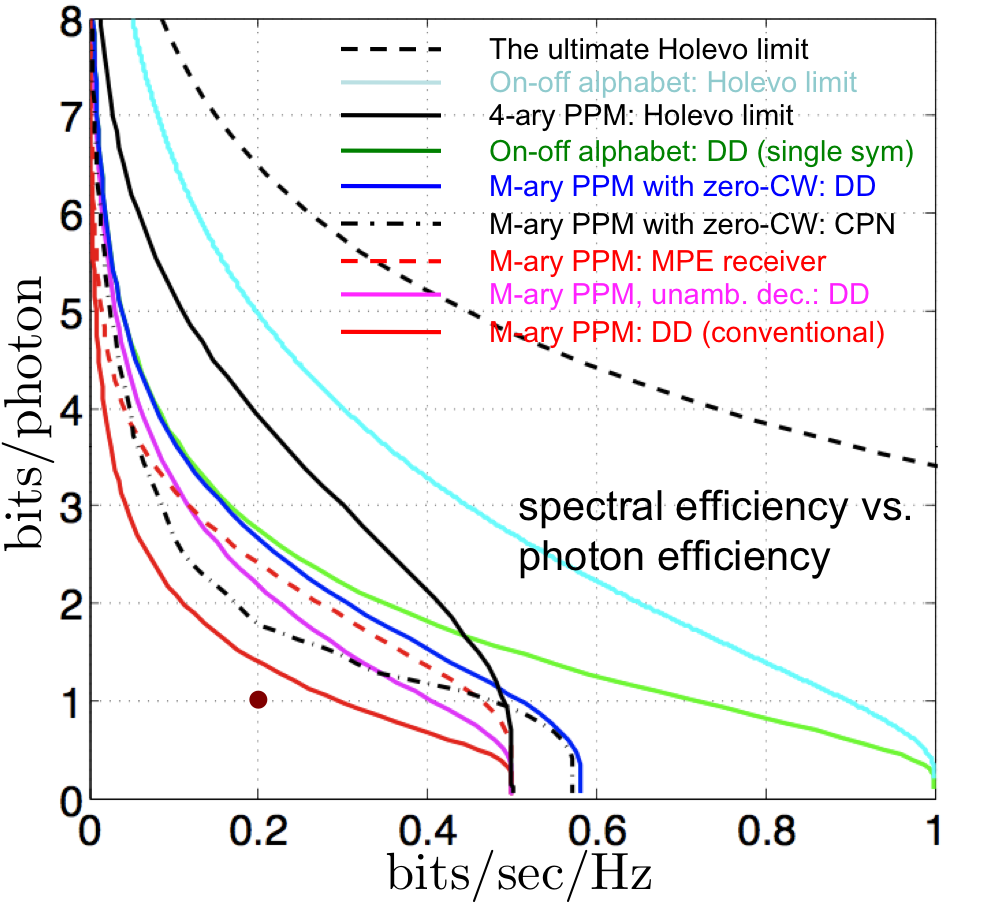}
\end{center}
\vspace{-10pt}
\caption{Spectral efficiency (bits/sec/Hz) versus energy efficiency (bits/photon) tradeoff, for a single-spatial-mode pure-loss optical channel. The black dashed line is the Holevo limit, which can't be exceeded by any modulation/receiver.}
\label{fig:capacities}
\end{figure}

In Fig.~\ref{fig:capacities}, we plot the spectral efficiency (bits/sec/Hz) versus the photon efficiency (bits encoded per received photon) for a single-spatial-mode pure-loss optical channel, such as a satellite-to-satellite link. PPM is routinely used for such communication links, whose performance is bounded by the red solid curve. The maroon dot below the red curve represents a practically very ambitious, albeit an achievable capacity using PPM with the conventional direct-detection (DD) receiver. At a pulse-repetition rate of $200$ MHz (${\sim}5$nsec pulses), and assuming a quasi-monochromatic $1.55$$\mu$m carrier, $1$ bpp and $0.2$ bits/sec/Hz translates to a data rate of $40$ Mbps at $5$ pW of power collected by the receiver aperture. The black dashed line is the ultimate Holevo bound, which is an (achievable) upper bound to the Shannon capacity of any modulation format and any receiver. In order to achieve the Holevo limit, one in general would need to make joint-detection measurements over long blocks of symbols. For instance, the cyan solid line is the Holevo limit for the on-off-keying (OOK) alphabet, with no restriction on the receiver measurement, whereas the green line is the Shannon limit of the OOK alphabet with the conventional DD receiver. Codebooks and structured receivers that can bridge this gap are not yet known. The PPM modulation symbols can be thought of as a codebook with an underlying OOK alphabet. Hence, the performance of PPM signaling will be upper bounded by the cyan line---the Holevo limit for OOK-DD. The black solid line is the Holevo limit of the $4$-ary PPM alphabet; though, a receiver than attains this capacity using $4$-ary PPM, {\em must} make soft decisions after each PPM symbol interval, and make joint measurements over many PPM codewords. The magenta solid line is the performance of the PPM alphabet with the DD receiver, when the erasure output is {\em not} mapped to one of the $M$ PPM words~\cite{Tak2010}, which can interestingly outperform the Helstrom minimum probability of error (MPE) measurement (red dashed plot), and also the exact-nulling CPN receiver (black dash-dotted plot), for high $M$, and in the low spectral efficiency high photon efficiency region. All the plots of PPM capacities (except the black solid line) have been optimized over $M \ge 4$. The black dash-dotted curve (CPN receiver) and the blue solid curve (DD) pertain to the PPM alphabet with one additional (all-zero) codeword, with an optimal prior probability assigned to the all-zero codeword. More results and an elaborate discussion on achievable and ultimate limits on information rates will be given in~\cite{Guh2010}.

The mathematical structure as well as the physical realizations of quantum-limited optical detection to achieve quantum limits to the hard-decoding minimum symbol error rates and the quantum limits to channel capacity, is an area ripe for research---both in theory and experiments. We hope that the above analysis would lead to useful insight towards structured codes and joint-detection receivers to approach the ultimate Holevo limits to the capacity of optical communications.

The authors thank Prof. Jeffrey H. Shapiro, MIT and Dr. Zachary Dutton, BBN, for several helpful discussions.


\begin{thebibliography}{10}
\bibitem{Bar1969} Bar-David, ``Communication under the Poisson regime," IEEE Trans. Information Theory, {\bf 15}, pp. 31Ð37, 1969. 
\bibitem{Hel1967} C. W. Helstrom, ``Detection theory and quantum mechanics," Information and Control, {\bf 10}, 1967. 
\bibitem{Gio2004} V. Giovannetti, S. Guha, S. Lloyd, L. Maccone, J. H. Shapiro, and H. P. Yuen, ``Classical capacity of lossy bosonic channels: the exact solution," Phys. Rev. Lett. {\bf 92,} 027902, 2004.
\bibitem{Sha2005} J. H. Shapiro, S. Guha, B. I. Erkmen, The J. of Opt. Net. (2005). 
\bibitem{Hel1976} C.W. Helstrom. ``Quantum Detection and Estimation Theory New York: Academic Press, 1976. 
\bibitem{Dol1973} S.J. Dolinar, Jr., ``An optimum receiver for the binary coherent state quantum channel," MIT Research Laboratory of Electronics Quarterly Progress Report 111, Massachussetts Institute of Technology, Cambridge, Massachussetts, pp. 115--120, October 1973. 
\bibitem{Coo2007} R.L. Cook, P.J. Martin, and J.M. Geremia, ``Optical coherent state discrimination using a closedÐloop quantum measurement," Nature, {\bf 446}, pp. 774--777, April 2007. 
\bibitem{Ken1973} R.S. Kennedy, ``A near-optimum receiver for the binary coherent state quantum channel," MIT Research Laboratory of Electronics, Quartely Progress Report 108, Cambridge, pp. 219--225, January 1973. 
\bibitem{Dol1983} S.J. Dolinar, Jr, ``A near-optimum receiver structure for the detection of M Ðary optical PPM signals," JPL TDA Prog. Rep., {\bf 42--72}, (1983). 
\bibitem{Bon1993} R.S. Bondurant, ``Near-quantum optimum receivers for the phase-quadrature coherent-state channel", Opt. Lett., {\bf 18}, 22, Nov. 15, 1993.
\bibitem{Bor2004} D. M. Boroson, A. Biswas, and B. L. Edwards, ``MLCD: Overview of NASA's Mars laser comm. demonstration system,Ó in SPIE Proc., Free Space Laser Comm. Tech. XVI, {\bf 5358}, pp. 16--23, (2004).
\bibitem{Sha1985} J. H Shapiro, IEEE J. Quantum Electron. {\bf QE-21,} 237--250 (1985).
\bibitem{Tak2008} M. Takeoka and M. Sasaki, ``Discrimination of the binary coherent signal: Gaussian-operation limit and simple non-Gaussian near-optimal receivers," Phys. Rev. A {\bf 78}, 022320 (2008).
\bibitem{Tak2010} A. Waseda, {\em personal communication} (2010).
\bibitem{Guh2010} S. Guha, J. H. Habif, and M. Takeoka, {\em to be submitted to JMO} (2010).
\end{thebibliography}
\end{document}